\documentclass{aa}

\usepackage{graphicx}
\usepackage{txfonts}
\usepackage[usenames,dvipsnames]{xcolor}
\newcommand{\hii}{\relax \ifmmode {\mbox H\,{\scshape ii}}\else H\,{\scshape ii}\fi}
\newcommand{\mi}{\relax \ifmmode {\mu{\mbox m}}\else $\mu$m\fi}
\newcommand{\ha}{\relax \ifmmode {\mbox H}\alpha\else H$\alpha$\fi}
\newcommand{\hb}{\relax \ifmmode {\mbox H}\beta\else H$\beta$\fi}

\newcommand{\sii}{\relax \ifmmode {\mbox S\,{\scshape ii}}\else S\,{\scshape ii}\fi}
\newcommand{\siii}{\relax \ifmmode {\mbox S\,{\scshape iii}}\else S\,{\scshape iii}\fi}
\newcommand{\siv}{\relax \ifmmode {\mbox S\,{\scshape iv}}\else S\,{\scshape iv}\fi}
\newcommand{\oiv}{\relax \ifmmode {\mbox O\,{\scshape iv}}\else O\,{\scshape iv}\fi}
\newcommand{\nii}{\relax \ifmmode {\mbox N\,{\scshape ii}}\else N\,{\scshape ii}\fi}
\newcommand{\niii}{\relax \ifmmode {\mbox N\,{\scshape iii}}\else N\,{\scshape iii}\fi}
\newcommand{\oi}{\relax \ifmmode {\mbox O\,{\scshape i}}\else O\,{\scshape i}\fi}
\newcommand{\oii}{\relax \ifmmode {\mbox O\,{\scshape ii}}\else O\,{\scshape ii}\fi}
\newcommand{\hei}{\relax \ifmmode {\mbox He\,{\scshape i}}\else He\,{\scshape i}\fi}
\newcommand{\heii}{\relax \ifmmode {\mbox He\,{\scshape ii}}\else He\,{\scshape ii}\fi}
\newcommand{\oiii}{\relax \ifmmode {\mbox O\,{\scshape iii}}\else O\,{\scshape iii}\fi}
\newcommand{\ariii}{\relax \ifmmode {\mbox Ar\,{\scshape iii}}\else Ar\,{\scshape iii}\fi}
\newcommand{\arv}{\relax \ifmmode {\mbox Ar\,{\scshape v}}\else Ar\,{\scshape v}\fi}
\newcommand{\arii}{\relax \ifmmode {\mbox Ar\,{\scshape ii}}\else Ar\,{\scshape ii}\fi}
\newcommand{\ariv}{\relax \ifmmode {\mbox Ar\,{\scshape iv}}\else Ar\,{\scshape iv}\fi}
\newcommand{\neii}{\relax \ifmmode {\mbox Ne\,{\scshape ii}}\else Ne\,{\scshape ii}\fi}
\newcommand{\neiii}{\relax \ifmmode {\mbox Ne\,{\scshape iii}}\else Ne\,{\scshape iii}\fi}
\newcommand{\nev}{\relax \ifmmode {\mbox Ne\,{\scshape v}}\else Ne\,{\scshape v}\fi}

\newcommand{\rdostres}{\relax \ifmmode {\,\mbox{R}}_{\rm 23}\else \,\mbox{R}$_{\rm 23}$\fi}

\newcommand{\ciii}{\relax \ifmmode {\mbox O\,{\scshape iii}}\else C\,{\scshape iii}\fi}
\newcommand{\civ}{\relax \ifmmode {\mbox O\,{\scshape iii}}\else C\,{\scshape iv}\fi}

\newcommand{\gsim}{\hbox{\rlap{\lower.55ex\hbox{$\sim$}} \kern-.3em
\raise.4ex \hbox{$>$}}}
\newcommand{\lsim}{\hbox{\rlap{\lower.55ex\hbox{$\sim$}} \kern-.3em
\raise.4ex \hbox{$<$}}}

\newcommand{\micron}{\hbox{$\mu$m}}

\begin{document}

   \title{Exploring the hardness of the ionizing radiation with the infrared softness diagram}
\subtitle{II. Bimodal distributions in both the ionizing continuum slope and the excitation in active galactic nuclei}

\titlerunning{The infrared softness diagram in AGN}

\author{
        E.~P\'{e}rez-Montero\inst{\ref{IAA}}  \and
        J.~A.~Fern\'andez-Ontiveros\inst{\ref{CEFCA},\ref{CEFCA-UA}} \and
        B.~P\'{e}rez-D\'{i}az\inst{\ref{IAA}}  \and
        J.~M. V\'{i}lchez\inst{\ref{IAA}} \and
        R. Amor\'in \inst{\ref{IAA},\ref{ARAID}}
}

\institute{
Instituto de Astrof\'{i}sica de Andaluc\'{i}a (CSIC), Apartado 3004, 18080 Granada, Spain \label{IAA}
\and
Centro de Estudios de Física del Cosmos de Aragón (CEFCA), Plaza San Juan 1, 44001 Teruel, Spain\label{CEFCA}
\and
Unidad Asociada CEFCA-IAA, CEFCA, Unidad Asociada al CSIC por el IAA, Plaza San Juan 1, 44001 Teruel, Spain\label{CEFCA-UA}
\and
ARAID Foundation, Centro de Estudios de F\'isica del Cosmos de Arag\'on (CEFCA), Unidad Asociada al CSIC, Plaza San Juan 1, 44001 Teruel, Spain\label{ARAID}
}

   \date{Received XXX; accepted YYY}


  \abstract
   {}
 {After exploring a version of the infrared (IR) softness diagram to characterize the hardness of the incident ionizing radiation in star-forming regions, we exploit the availability of high-excitation lines in the same spectral regime to explore its potential use for studying the narrow-line regions of active galactic nuclei (AGNs).}
 {We adapted the IR softness diagram to consider very high-excitation lines, such as [\nev]$14.3,24.3\, \rm{\micron}$ or [\oiv]$25.9\, \rm{\micron}$. The measured emission-line ratios were included as inputs for the code {\sc HCm-Teff-IR} in order to provide a quantification for the ionization and the $\alpha_{\rm OX}$. The latter is sensitive to the spectral shape of the incident continuum in AGNs. We applied this code to a large AGN sample that included different spectral types with available {\it Spitzer}/IRS, {\it Herschel}/PACS, and/or SOFIA/FIFI-LS mid-IR spectroscopic observations.}
 {The combination of the ([\neii]$12.8\, \rm{\micron}$+[\neiii]$15.6\, \rm{\micron}$)/[\nev]$14.3\, \rm{\micron}$ and [\oiii]$52,88\, \rm{\micron}$/[\oiv]$25.9\, \rm{\micron}$ emission line ratios is a robust proxy for the shape of the ionizing continuum in AGNs. Given the difficulties to observe the [\oiii]$52,88\, \rm{\micron}$ lines, an alternative based on the [\siii]$18.7\, \rm{\micron}$+[\siv]$10.5\, \rm{\micron}$ lines can be used. The inclusion of very high-excitation lines in the input ratios for {\sc HCm-Teff-IR} to derive both $\alpha_{\rm OX}$ and $U$ for the studied sample points to a bimodal distribution of galaxies. One of the peaks is characterized by relatively harder values of $\alpha_{\rm OX}$ around -1.4 in combination with low values for log $U$ around -2.4, while the other peak shows a softer $\alpha_{\rm OX} \sim $ -1.7 and high values of log $U$ $\sim$-1.5. This result is consistent with the existence of two very distinct AGN populations, where one has a softer ionizing continuum, possibly dominated by a radiatively efficient accretion disk in bright Seyfert nuclei. In contrast, we observed a harder radiation field in low-luminosity AGNs, where the accretion disk is expected to recede.}   
  {}


   \maketitle
%

\section{Introduction}

The study and characterization of the nature of ionizing sources in galaxies has always been a recurrent area of research given the diverse possibilities that can lead to the production of an emission-line spectrum typical of regions dominated by the emission from ionized gas. These can include episodes of massive star formation, stellar populations dominated by the presence of hot low-mass evolved stars (HOLMES), shocks, or active galactic nuclei (AGNs).

Emission lines can provide valuable information on the gas properties and the sources responsible for the excitation and ionization of the elements producing them. In this context, given the observational difficulties found in directly studying the spectral regimes where the possible different sources can be distinguished (mainly X-rays and ultraviolet), other diagnostics based on the relative fluxes of the lines emitted by the gas can be used instead, as they include strong recombination lines (RLs) and collisionally excited lines (CELs), which are more accessible since they are observed in the optical and the infrared (IR) ranges.

The gas velocity dispersion derived from line widths provides important evidence of the mechanical energy supplied by different sources and thus can be used to derive their nature. However, when sufficient spectral resolution is not available, alternative methods based on the line fluxes are required. In this case, there is a proliferation of diagnostic diagrams based on the flux of certain CELs relative to the strongest RLs, such as H$\alpha$ or H$\beta$. These include the well-known Baldwin, Phillips, and Terlevich diagrams \citep[BPT;][]{bpt}, both empirically \citep[e.g.,][]{kauffmann03} and theoretically \citep[e.g.,][]{kewley06} calibrated, which are widely used to elucidate the nature of the ionizing source. On the other hand, as these diagnostic diagrams are based only on certain optical line fluxes, they are insufficient to produce a precise analysis of the incident spectral energy distribution (SED) and are most often used to establish a first-order selection in large samples prior to another kind of analysis, such as the determination of chemical abundances from these same emission lines. A more sophisticated approach to describing the nature and shape of the ionizing continuum in AGNs is based on the analysis of several emission line ratios, both in the optical \citep{baldwin75,shields78,malkan82} and the IR range \citep{alexander99,alexander00,melendez11}. Nevertheless, most of these efforts are based on deterministic analyses using a relatively complex set of models, leading to ambiguous results on the shape of the ionizing continuum.

An alternative approach to characterize the hardness of the radiation field in star-forming regions is the softness diagram \citep{pmv09,kumari21}, which is based on the flux ratios of emission lines from ionic species of consecutive ionization stages. This method leverages the dependence of the so-called softness parameter ($\eta$), defined by \cite{vp88}, on the hardness of the ionizing radiation. In a previous paper of this series (\citealt{pm24}, hereinafter Paper I), we analyzed a version of this diagram exclusively based on mid-IR lines, taking advantage of the large amount of data on emission-line galaxies available in this regime, such as the IDEOS catalog \citep{Hernan-Caballero_2016,Spoon_2022} based on {\it Spitzer}/IRS, but also considering the upcoming new data from JWST.

Among the various advantages that IR lines present compared to optical ones is their significantly lower dependence on extinction, which makes them especially appropriate for penetrating the large optical depths in galaxy dusty regions, particularly in deeply obscured objects such as ultra-luminous infrared galaxies (ULIRGs). Another major advantage of IR transitions is that their emissivities depend only marginally on the electron temperature because the associated energy levels are much closer to the ground state. Additionally, this spectral regime presents transitions from highly ionized species that trace a more energetic spectral regime in the incident SED. This can reduce the uncertainty in the derivations affecting the optical tracers due to the contribution from low-excitation emission lines, such as [\sii]$\lambda \lambda 6716, 6731$ \citep{pm23}.

Several authors have proposed alternative $\eta$ parameters  based on IR emission lines from Ar, Ne, or S \citep{martin02,morisset04,pmv09}, including the following expression examined in Paper I:
\begin{equation}
\eta'_{\rm IR} = \frac{\rm Ne2Ne3}{\rm S3S4} = \frac{{\rm [NeII]}_{12.8 \mu{\rm m}} / {\rm [NeIII]}_{15.6 \mu{\rm m}}}{{\rm [SIII]}_{18.7 \mu{\rm m}} / {\rm [SIV]}_{10.5 \mu{\rm m}}}.
\end{equation}

In Paper I, we also presented an adaptation of the {\sc HCm-Teff} code \citep{hcm-teff}, originally designed to determine the equivalent effective temperature ($T_*$) and the ionization parameter ($U$), for use with IR transitions. This approach is based on a Bayesian-like comparison between the predictions of a comprehensive photoionization model grid and the appropriate IR emission-line ratios observed, as included in the definition of the softness diagram based on the $\eta'_{\rm IR}$ parameter. This step is necessary to break the simultaneous dependence of $\eta'_{\rm IR}$ on $T_*$, $U$, and metallicity ($Z$). When applied to the IDEOS sample, our results confirmed that no additional sources harder than typical massive stars were required to reproduce the emission-line ratios observed in all star-forming galaxies, including ULIRGs.

\begin{figure}
   \centering
\includegraphics[width=\columnwidth,clip=]{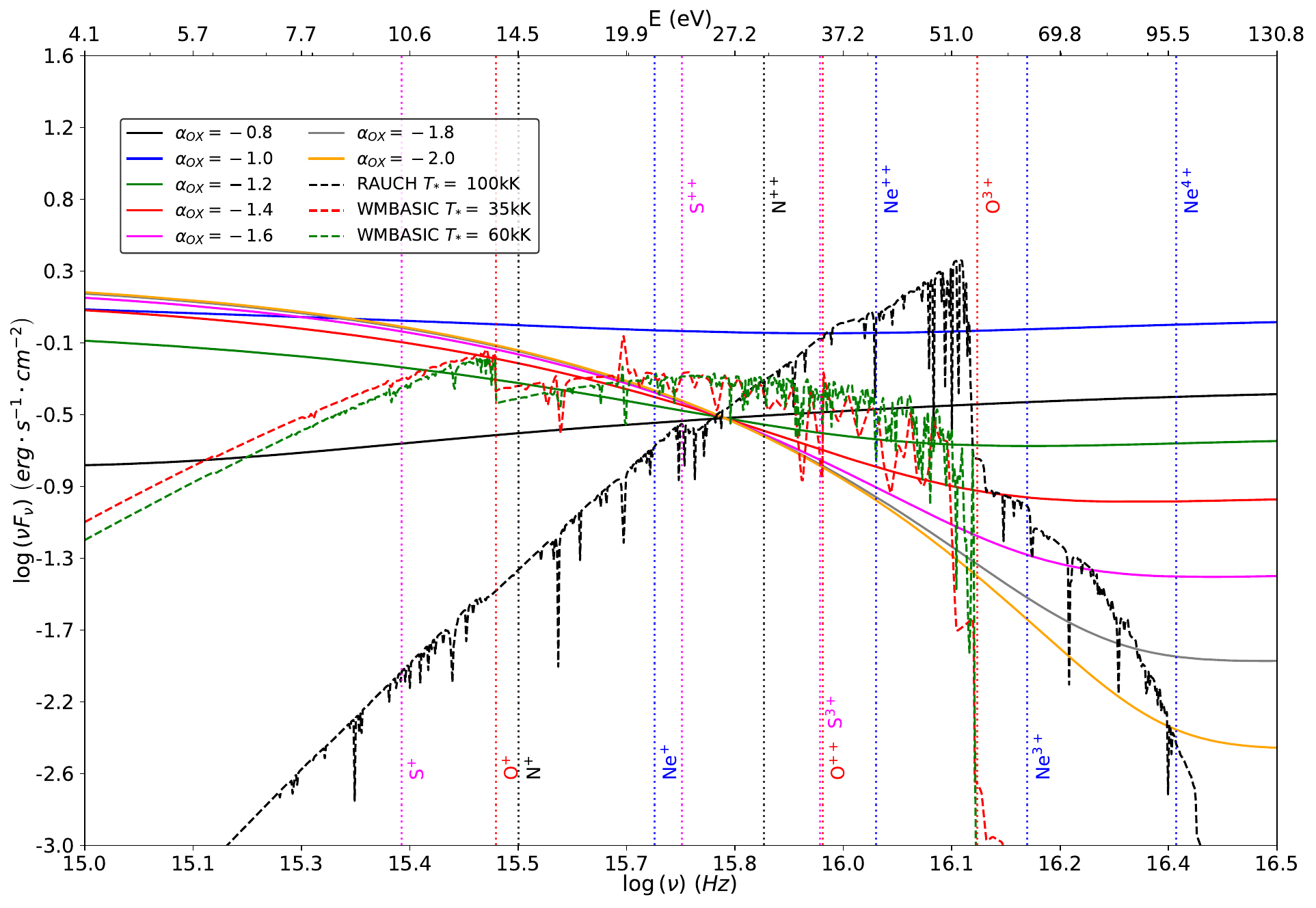}
\caption{Normalized SEDs of some of the sources considered by the photoionization models used by {\sc HCm-Teff} for both AGNs (solid lines) and star-forming regions (dashed lines), including individual massive stars from WM-Basic \citep{wmbasic} and planetary nebulae \citep{rauch}. The ionization potentials of the ions whose optical and IR emission lines are used in our calculations are indicated as vertical dotted lines.}
\label{SEDs}%
\end{figure}

In this work, we propose the study of softness parameters and diagrams based on IR lines in order to examine the shape of the incident radiation field in AGNs. Given the much harder SED expected in these objects, it is known that the more widely low-excitation lines in the optical are ineffective for this aim, so we explored the same parameter in the IR, assessing the use of other high-excitation emission lines in this range, such as [\nev]14.3, 24.3 \micron\ or [\oiv]25.9 \micron, whose presence has been predicted \citep{spinoglio92} and extensively confirmed in the literature for AGNs \citep[e.g.,][]{genzel98,sturm02,melendez08,armus23}.

As in the case of $T_*$ for star-forming regions, we consider the index $\alpha_{\rm OX}$ as a parameter that probes the shape of the incident SED in AGNs. This index represents the slope of the power-law continuum between 2 keV and 2500 \AA, and it can be used to parametrize the hardness of the SED in each model.
A high $\alpha_{\rm OX}$ value indicates relatively strong X-ray emission compared to the UV range, which can highlight differences in the physics of the environment near the supermassive black hole \citep[e.g.,][]{Tananbaum79, Vignali03}. Figure~\ref{SEDs} shows the effect of changing the $\alpha_{\rm OX}$ value in different double-peaked power laws compared with the energy of the ions whose emission lines are typically observed in the mid-IR range as well as with other SEDs typical of massive stars.

The paper is organized as follows. In Sect. 2, we describe the observational samples of compiled mid- and far-IR emission lines in the sample of AGNs. In Sect. 3, we describe the grids of photoionization models used to derive emission-line fluxes in the IR and build the different versions of the softness diagram. The most suitable form of this diagram for studying AGNs is discussed in Sect. 4, which is accompanied by a description of the adaptation of the {\sc HCm-Teff} code for the IR range. In Sect. 5, we present and discuss our results from the application of this code to the different samples of objects. Finally, in Sect. 6, we summarize our results and present our conclusions.

\section{Data samples}
In this work, we compiled narrow nebular emission-line fluxes spectroscopically measured in the mid-IR in AGNs by combining catalogs from \textit{Spitzer}, \textit{Herschel}, Akari, and SOFIA. The data were mainly obtained from three sources: 1) the Infrared Database of Extragalactic Observables\footnote{\url{http://ideos.astro.cornell.edu}} \citep[IDEOS][]{Hernan-Caballero_2016,Spoon_2022} from {\it Spitzer} archive; 2) the AGN  sample from \citet{jafo16} and \citet{Spinoglio_2022} combining {\it Spitzer}, {\it Herschel} and SOFIA; 3) the (U)LIRG catalog from \citet{Imanishi_2010} observed with {\it Akari}.

While the last two samples present an spectral classification of their AGN-dominated nature using both the optical (through BPT diagnostics) and IR ranges (through the presence of [\nev] and [\oiv] lines), this is not the case for the former one. In order to obtain a sample of AGN-dominated galaxies in the IDEOS sample, we used the strength (in terms of equivalent width; EW) of the polycyclic aromatic hydrocarbon (PAH) emission at 6.2 $\mu$m, which is expected to discriminate between star-formation and AGN activity \citep{puget89,armus07,hernan20}. The selected sample verifies a EW(PAH\ensuremath{_{6.2\micron}) \leq} $0.06\,\mu$m, which is typical of galaxies dominated by AGN activity \citep{armus07}. Additionally, we required the detection of higher ionic species. Specifically, the selected sources have both [\nev], either at 14$\mu$m or 24$\mu$m, and [\oiv] detections in the IDEOS catalog, which are indicative of AGN activity \citep{armus07}. The subsequent analysis requires in addition [\neii] at 12.8\,$\mu$m, [\neiii] at 15.6\,$\mu$m, and either [\oiii] at 52\,$\mu$m or 88\,$\mu$m, or [\siv] at 10.5\,$\mu$m along with [\siii] at either 18.7\,$\mu$m or 33.5\,$\mu$m.
Only emission line fluxes with a minimum signal-to-noise ratio (S/N) of 3 were used.
After this selection to ensure the AGN-dominated nature of the selected objects, our final sample consists of  484 objects, considering as different objects different measurements taken at a distinct wavelength regime, spatial coverage or spectral resolution of the same galaxy. In any case, the number of duplicated objects (13) is very low and it does not  affect the statistical significance of our results.

As shown in Paper I for star-forming objects, the chemical content of the gas may be a relevant input parameter for the determination of the hardness using emission lines. Thus, we derived the total oxygen abundance in a small subsample of the resulting compiled AGNs in order to better constrain the models used to characterize them. The determination of the oxygen abundance was done, as described in \cite{pd24}, using a version of the {\sc HCm-IR} code \citep{jafo21b} adapted for its use for AGNs \citep{pd22}. As described in \citet{pd22}, the code only provides a solution for the derivation of the oxygen abundance of the gas-phase in AGNs if at least one hydrogen recombination line is supplied. For this reason, available measurements of the hydrogen recombination line Brackett \ensuremath{\alpha} {\it Akari/IRC} (2.5-5 $\mu$m, \citealt{Imanishi_2010}), were added for some objects in the IDEOS catalog.

\section{Description of the photoionization models}

Several different grids of photoionization models were computed in order to be compared with the compiled observations, and provide a complete model database for the code {\sc HCm-Teff} to obtain an estimate of the hardness of the incident radiation field in AGNs.

To this aim, we calculated grids of models similar to those described in \cite{hcm-agn} and \cite{pd22}, used for the calculation of chemical abundances in the narrow-line region (NLR) of AGNs by {\sc HCm}. These models use a composite SED as incident source, one component representing the Big Blue Bump peaking at 1 Ryd, and the other a power law with spectral index $\alpha_x$ = -1, representing the non-thermal X-ray radiation in the 13.6\,eV to 100\,keV range \citep{elvis94,korista97,melendez11}. As free parameter in the grids we computed models varying the values of the spectral index $\alpha_{\rm OX}$ in the -2.0 to -0.8 range with a 0.2 step. Typical measurements of $\alpha_{\rm OX}$ vary depending on the sample and the methods used. For instance, \cite{miller11} conducted a comprehensive study of $\alpha_{\rm OX}$ in a large sample of AGNs, finding values typically ranging from -1.5 to -1.1, depending on luminosity and redshift.

{\sc HCm-IR} and {\sc HCm-Teff} use models computed with {\sc Cloudy} v17.01\footnote{No significant differences were found between these models and test models run with {\sc Cloudy} v23.01.} \citep{cloudy}. The models span a range in total oxygen abundance from 12+log(O/H) = 8.1 to 9.0 in bins of 0.3 units, and log $U$ from -4.0 to -0.5 in bins of 0.25 units. These ranges cover the values derived in the different AGN samples analyzed using the code {\sc HCm} (e.g., \citealt{hcm-agn,dors20,pd21,pd22,pm23,pd24}).

All other chemical abundances were scaled to the solar proportions as described by \cite{asplund09}, with the exception of N, which follows the empirical relation derived by \cite{pm14}. All models were calculated assuming a plane-parallel geometry, assuming a filling factor of 0.1 and a constant density of 500 cm$^{-3}$, typical for these objects and below the critical density of the emission lines involved in this study. A more thorough discussion on the effects of density in this models can be found in \cite{pd22,pd24}. Two possible scenarios about the presence of dust grains mixed with the gas were considered in all models: one with a standard Milky Way proportion, and another way without dust grains in order to check the impact of the presence of dust in the results. In those models with dust, we did not consider depletion factors in the chemical abundances, so we always refer to the values in the gas-phase. In addition, the stopping criterion of each model is changed to consider either 99.9\%, 98\% or 2\% of free electrons to explore how considering a matter-bounded geometry alters the emission-line ratios between low- and high-excitation lines. All these possibilities yield a total of 1\,872 models in the final grid.

\begin{figure*}
   \centering
\includegraphics[width=0.33\textwidth,clip=]{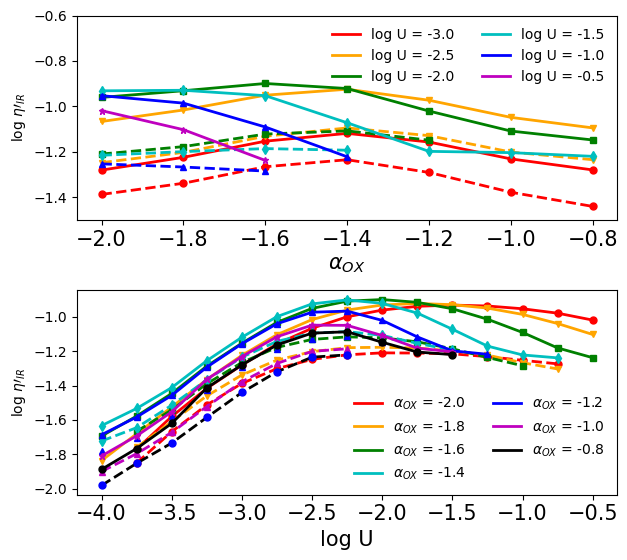}
\includegraphics[width=0.33\textwidth,clip=]{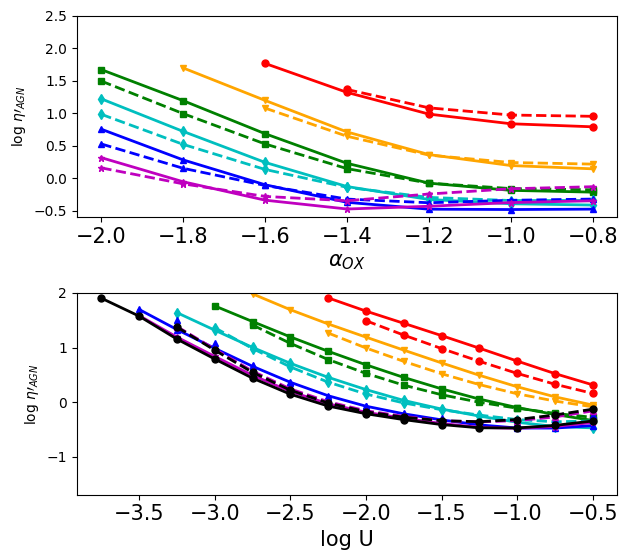}
\includegraphics[width=0.33\textwidth,clip=]{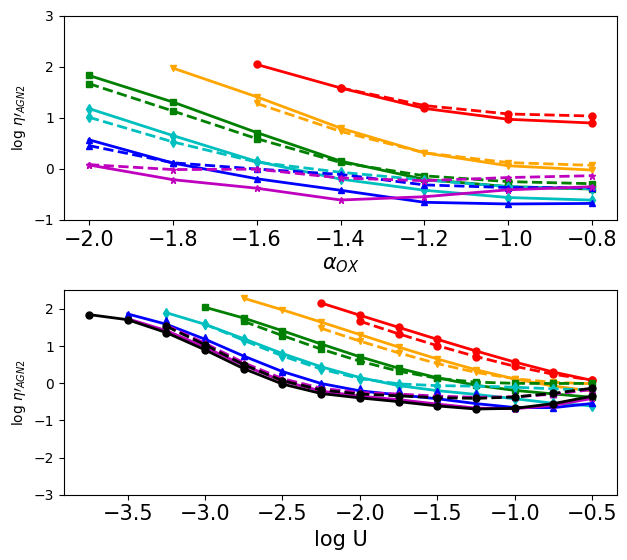}

\caption{Variation predicted by different sequences of photoionization models of log $\eta\prime_{\rm IR}$ (left panels), log $\eta\prime_{\rm AGN}$ (middle panels), and log $\eta\prime_{\rm AGN2}$ (right panels) as a function of the parameter $\alpha_{\rm OX}$ (upper panels) and log $U$ (lower panels). The models represented here were calculated with a stopping criterion when the fraction of free electrons is 98\%. In all panels  the solid lines represent models calculated with 12+log(O/H) = 9.0, while dashed lines assume a value of 8.1.}
\label{eta-mod}%
\end{figure*}

\section{Determining the hardness in AGNs from IR lines}

\subsection{A softness parameter in the mid-IR valid for AGNs}

Given the much harder SED assumed to be the cause of the ionization of the elements in the NLR of AGNs, as  compared to massive stars (see Fig.~\ref{SEDs}), we first explored to what extent the above-defined $\eta'_{\rm IR}$ can be useful in tracing the hardness of the radiation field in these objects.

To do so, we inspected the sequences of photoionization models described in the above subsection and obtained the values for the log $\eta'_{\rm IR}$ parameter as a function of $\alpha_{\rm OX}$ and log $U$, represented in Fig.~\ref{eta-mod}. In the left panels of the figure, we show the variation of log $\eta'_{\rm IR}$ predicted by the models as a function of $\alpha_{\rm OX}$ (upper panel) and log $U$ (lower panel) for two different values of metallicity (i.e., 12+log(O/H) = 9.0, 8.1). The models shown in the figure assume a stopping criterion corresponding to a fraction of free electrons of 98\%.
Considering that higher values of $\alpha_{\rm OX}$ correspond to harder SEDs, as can easily be verified in Fig.~\ref{SEDs}, no clear correlation can be seen between the value of $\eta'_{\rm IR}$, defined for its use in star-forming galaxies, and the hardness of the radiation field in AGNs. This is due to the relatively low ionization potential of the lines involved in $\eta'_{\rm IR}$, which mostly probe the continuum of the big blue bump below 3 Ryd and therefore respond poorly to the power-law continuum at higher energies.

Nonetheless, we can take advantage of the fact that in the mid-IR range we can measure higher ionization levels of the same species, which can help us extend the energetic range of the incident SED that we can assess using the available emission lines \citep[e.g.,][]{spinoglio15,jafo16}. This is more helpful in distinguishing possible differences in the power-law. As an alternative measurable in the mid- to far-IR spectral range, we defined
\begin{equation}
\eta'_{\rm AGN} = \frac{\rm Ne23Ne5}{\rm O3O4} = \frac{({\rm [NeII]}_{12.8 \mu{\rm m}}+{\rm [NeIII]}_{15.6 \mu{\rm m}})/{\rm [NeV]}_{14.3 \mu{\rm m}}}{{\rm [OIII]}_{52 \mu{\rm m}}/{\rm [OIV]}_{25.9 \mu{\rm m}}}.
\end{equation}

This modification allowed us to largely extend the energetic range evaluated by the parameter, covering wavelengths where the possible differences in the power-law index defining the incident radiation field in AGNs can be properly distinguished. We note that this parameter can also be defined using alternative equivalent transitions for the same ions in a similar spectral range, as in the case for [\nev] at 24.3 \micron, and [\oiii] at 88 \micron, which can replace the transitions defined in the above expression.

In the middle panels of Fig.~\ref{eta-mod}, we show the variations of this new parameter $\eta'_{\rm AGN}$ as a function of $\alpha_{\rm OX}$ and log $U$. We note that this parameter is only shown when the models predict a certain level of emission for all involved lines, so it cannot be defined for certain low $\alpha_{\rm ox}$ and $U$ where the models do not predict any emission of the high-excitation lines. As can be seen, the parameter now clearly traces the hardness of the power-law, as it has lower values for higher values of $\alpha_{\rm OX}$, consistent with the original optical softness parameter defined by \cite{vp88}, which decreases for harder radiation fields. Another important characteristic of this new parameter is its very low dependence on metallicity, given that the assumed AGN SEDs have not got any variation as a function of it, added to the relatively much lower dependence of the mid-IR emission-line emissivities on electron temperature, as compared to optical lines.

On the contrary, one important caveat of this new form of the softness parameter is the fact that the far-IR [\oiii], either at 52 \micron\ or 88 \micron, are not covered at low redshift using the available facilities, once the SOFIA observatory has been retired, limiting the number of objects for which an analysis based on this parameter can be carried out. An alternative could be to replace this line using another emission line at a shorter wavelengths, more easily covered by mid-IR observations, corresponding to a similar energy range. One possibility could be the use of the sulphur emission lines of [\siv] at 10.5 \micron\ and [\siii] at 18.7 \micron, covering a very similar energy range (Fig.~\ref{SEDs}), although with a possible additional uncertainty coming from variations in the S/O abundance ratio \citep[e.g.,][]{pd24}. In this case, the alternative form of the softness parameter would be
\begin{equation}
\eta'_{\rm AGN2} = \frac{\rm Ne23Ne5}{\rm S34O4} = \frac{({\rm [NeII]}_{12.8 \mu{\rm m}}+{\rm [NeIII]}_{15.6 \mu{\rm m}})/{\rm [NeV]}_{14.3 \mu{\rm m}}}{({\rm [SIII]}_{18.7 \mu{\rm m}}+{\rm [SIV]}_{10.5 \mu{\rm m}})/{\rm [OIV]}_{25.9 \mu{\rm m}}}.
\end{equation}

As in the previous case, similar expressions can be explored using the [\siii] line at 33.5 \micron\ instead.
In the right panels of Fig.~\ref{eta-mod} we show the dependence of this parameter on $\alpha_{ox}$ and log $U$, where a very similar behavior to that obtained for the version of the parameter based on [\oiii] is seen, confirming its potential use to trace the hardness of the ionizing SED in AGNs.

\begin{figure*}
   \centering
\includegraphics[width=0.4\textwidth,clip=]{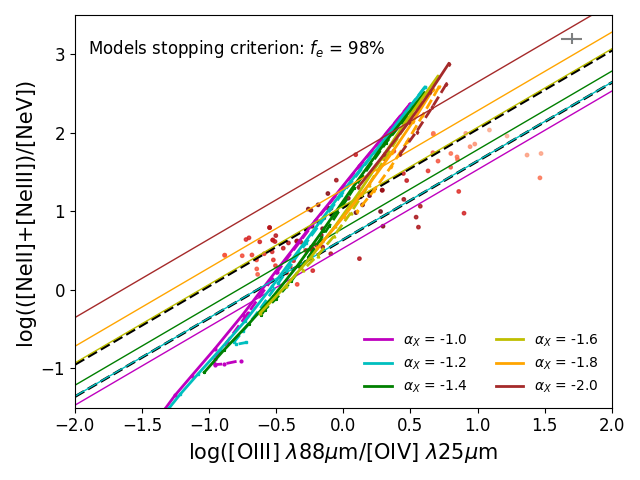}
\includegraphics[width=0.4\textwidth,clip=]{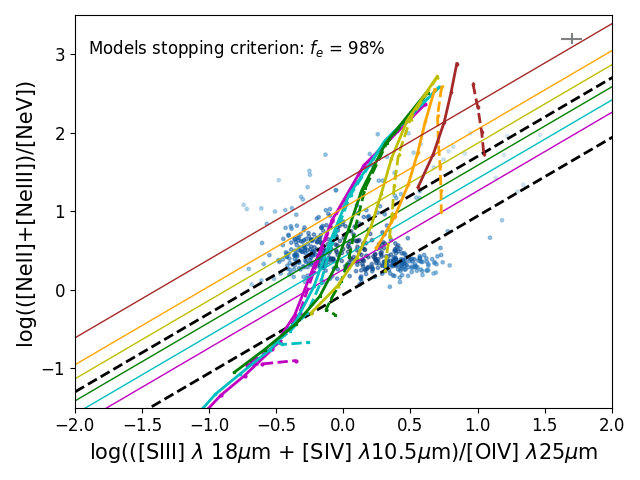}
\includegraphics[width=0.4\textwidth,clip=]{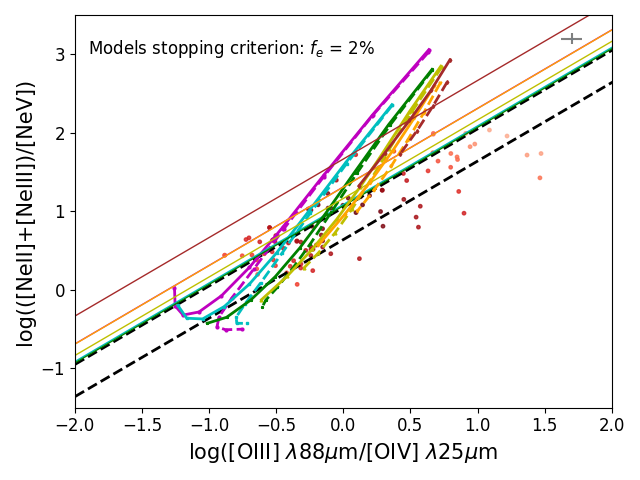}
\includegraphics[width=0.4\textwidth,clip=]{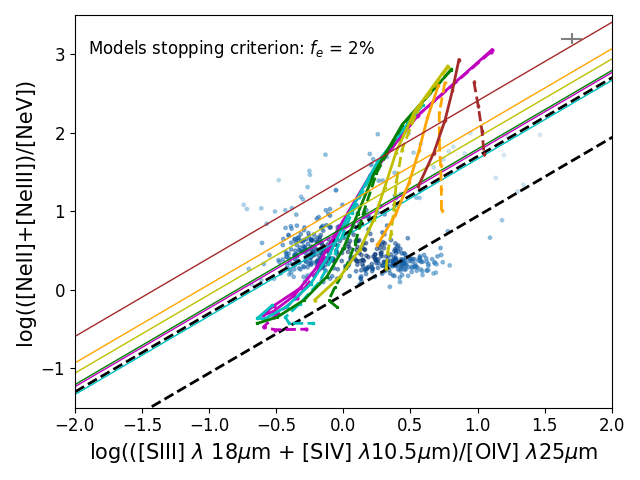}

\caption{Two versions of the softness diagram based on IR lines for NLR in AGNs: ([\neii]+[\neiii])/[\nev] in relation to [\oiii]/[\oiv] (left panels) and to ([\siii]+[\siv])/[\oiv] (right panels) for the compiled sample, represented as density points (red for the sample with [\oiii] and blue for [\siii]+[\siv]), as compared with sequences of photoionization models. In all panels the thick colored lines represent models calculated for different values of $\alpha_{\rm OX}$ and connecting different values for log $U$ from -0.5 in the lower left part of each sequence in steps of 0.25 dex. In the upper panels the stopping criterion of the models is that the fraction of free electrons is   98\%, while in the lower panels is 2\%. Solid lines in all panels represent models with dust grains, while dashed lines represent models without dust. The dashed black lines represent the average log $\eta\prime$ values obtained for the bimodal distributions obtained as described in Sect. 4.  The mean $\eta\prime$ values for each model sequence are also represented as solid thin lines of the same color (only shown for models with dust grains). In the upper right part of all panels, the cross indicates the typical observational errors for the represented sample.}
\label{eta-agn}%
\end{figure*}

In Fig.~\ref{eta-agn}, we compare the two above-defined parameters represented in two dimensions under the form of diagrams, in order to compare in the same space the observations described in Sect. 2, represented as density points, with the predictions from some of the photoionization models described in Sect. 3. As a first evident impression, the use of the mid-IR lines of [\siii] and [\siv] replacing the far-IR [\oiii] results very convenient, given the scarcity of points observed in that regime. In contrast, the IDEOS catalog allowed us to explore a significantly larger sample. As support of this, in our overall sample of 484 objects, we only found 97 presenting a measurement of the far-IR [\oiii] line, either at 52 $\mu$m or 88 $\mu$m, while [\siii]+[\siv], either with [\siii] at 18.7\,$\mu$m or 33.5\,$\mu$m, can be measured in 480 objects of the compiled sample.

The compiled objects shown in Fig.~\ref{eta-agn} are compared with the predictions in the same plane from several of the calculated photoionization models to show the trends as a function of the model input parameters. In all panels, sequences with the same colors and symbols join models of solar metallicity with the same $\alpha_{\rm OX}$, in the -2.0 to -1.0 range, connecting models with different log $U$ covering a range up to -0.5. Thin solid lines represent the mean log$\eta\prime$ for each sequence. The upper panels show the models calculated for a stopping criterion of 98\% free electrons, while this is of 2\% for the models shown in the lower panels. The solid lines show models calculated assuming dust mixed with the gas, while dashed lines show the same models when no dust is assumed.

As it can be seen, although the sequences of larger $\alpha_{\rm OX}$ cover the region of the diagram corresponding to high values for both ratios, they consistently show lower $\eta'_{\rm AGN}$ values (i.e., the lines of slope 1 in the diagram), while the models with lower $\alpha_{\rm OX}$ are situated in agreement with the expected higher values of $\eta'_{\rm AGN}$. Regarding $U$, the values go in the same direction as already observed in the version of the softness diagram studied for star-forming regions: higher values of $U$ are situated in the lower left part of each sequence (i.e., corresponding to low values of Ne23Ne5 and O3O4), while lower values of $U$ are situated in the upper right part (i.e., for high values of Ne23Ne5 and O3O4). Both diagrams hence show that they could in principle be used to determine both $\alpha_{\rm OX}$ and $U$.

However, the comparison of the different coverage of the objects by the models with different geometry in upper and lower panels of the same figure reveals that the assumptions taken for the models can be critical for the final derivation. As it can be seen in the upper panels, the models calculated at a fraction of 98\% better reproduce the distribution of observed $\eta\prime$ values, whereas the mean values are not well covered by the models assuming a 2\% of free electrons as a stopping criterion (lower panels). This suggests that possibly most of the studied objects have a matter-bounded geometry as opposed to a radiation-bounded geometry, leading to a fraction of escaping photons. In any case, we considered both geometry assumptions for our calculations.

The equivalent model sequences calculated without dust grains are consistent with studies by several authors who omit them to match the observations of certain UV lines (e.g., \citealt{nagao06,dors19}). Apart from a slight difference at lower values of $U$, this change does not appear to produce differences larger than those caused by other factors, such as geometry (which we confirm in the subsections below). The small differences found in this case are probably motivated by the much lower dependence of the IR lines on dust absorption. This again underlines the advantage of using IR lines for this kind of analysis, as they reduce the uncertainty due to extinction correction in the used emission lines.

\subsection{The {\sc Hii-Chi-mistry-Teff-IR code for AGNs}}\label{HCmTeff}

The use of the softness parameter based on IR lines, as defined in Eqs. 2 and 3, or using alternative ratios of emission-line fluxes involving low- and high-excitation states of ionization, can serve as a proxy for the hardness of the incident radiation. However, as discussed in \cite{hcm-teff} for the optical version of this parameter based on [\oii], [\oiii], [\sii], and [\siii], the $\eta\prime$ parameter presents additional dependencies on other functional parameters that can lead to incorrect interpretations for objects presenting different $\eta\prime$ values but also with very different metallicities or excitation levels. A similar behavior was found and studied in Paper I when the corresponding version of the parameter for the mid-IR was examined.

To address these issues, the code {\sc HCm-Teff-IR}\footnote{All versions of the code {\sc HCm} can be retrieved from the webpage \url{http://home.iaa.es/~epm/HII-CHI-mistry.html}} was developed by \cite{hcm-teff} for the case of optical lines, and later adapted to the IR in Paper I, to find more accurate solutions for $T_*$ by performing a Bayesian-like comparison between the corresponding observed emission-line ratios and the grid of photoionization models for star-forming regions.

Similarly, although the mean $\eta\prime_{\rm AGN}$ of the different model sequences correlate with their $\alpha_{\rm OX}$, there are also deviations as a function of $U$, as can be seen in Fig.~\ref{eta-agn}, and in lesser terms on $Z$, as can be seen in Fig.~\ref{eta-mod}, that must be analyzed in order to extract a more precise estimation of the hardness of the radiation field. Thus, the code serves as a tool to minimize these additional dependencies for the analysis of the incident radiation field hardness. Here, we therefore describe the adaptation of the code {\sc HCm-Teff-IR} so that it includes the emission lines predicted by the photoionization model grids shown in Sect. 3. Additionally, we discuss the appropriate emission-line ratios needed to construct the $\eta\prime_{\rm AGN}$ parameter in its various forms, adapted to the available observational data.

Once the parameters of the models that will be used (i.e., $f_e$, dust) are chosen, the code first interpolates the tables for the corresponding metallicity, which is treated as an input parameter, expressed as the total oxygen abundance. As in previous versions, if the metallicity is not given as input, the code considers all possible values, implying an enhanced uncertainty. The code then begins an iterative process along all the grid of models to create a $\chi^2$-weighted distribution of $\alpha_{\rm OX}$ and log $U$, whose mean and standard deviation are the found solutions. Notice that this is not the classical Bayesian approach in which a unique model with the least resulting $\chi^2$ is chosen, as this last procedure would be very inefficient to figure out the real uncertainty of the result. The corresponding $\chi^2$ values are obtained by the code as the sum of the quadratic differences between the observed emission-line ratios and those predicted by the models, which can be calculated with the emission-line fluxes given as input. For this version of the code, we ruled out in the grids those models corresponding to $\alpha_{\rm OX} = -0.8$ and log $U > -1.0$, because these do not follow any clear trend useful to derive these parameters from the chosen emission-line ratios, as can be seen in Fig.~\ref{eta-mod}.

In addition to the uncertainty associated with the standard deviation of the $\chi^2$-weighted distribution, the code also considers as a source of uncertainty the dispersion of the solutions found after a Monte Carlo iteration using as input values the nominal fluxes perturbed with the observational given errors.

The code admits as input the reddening-corrected fluxes in arbitrary units of the emission lines of the ratios involved in the several forms of the $\eta\prime_{\rm AGN}$ parameter, including:\footnote{Additionally, other emission lines such as [\arii] 7.0 \micron\ + [\ariii] 8.9 \micron\ and/or 21.8 \micron\ are also accepted by the code when they are given in relation to other high-excitation lines as [\arv] 7.9 \micron\ or 13.1 \micron. We did not find any object in the compiled samples with these lines, but their interest does not present any doubt considering their potential detection in many targets observed by JWST in the near future.}

\begin{itemize}
\item {[\neii]} 12.8\,\micron, [\neiii] 15.6\,\micron, and [\nev] 14.3 and/or 24.3\,\micron.
\item {[\siii]} 18.7 and/or 33.5\,\micron, and [\siv] 10.5\,\micron.
\item {[\oiii]} 52 and/or 88\,\micron, and [\oiv] 25.9\,\micron.
\end{itemize}

Specifically for the case of the mid-IR lines in this study, we did not apply any extinction correction due to the relatively low extinction values involved. Additionally, the narrow-line emission in AGNs is produced by an extended and diffuse gas region, which is les affected by the nuclear dust obscuration.
The code only considers an emission-line flux for the calculation when it can be used in combination with the emission-line flux for a complementary ionization stage (e.g., [\nev] is only used if, in addition, [\neii] and [\neiii] are also given). In the case where more than one line is given for a certain ion (e.g., [\nev], [\siii]), the code uses all forms of the corresponding ratios with equal weight. Therefore, it is up to the user to decide which line(s) to include if all of them are not going to be used.

\section{Results and discussion}

\subsection{The distribution of $\eta\prime$ for AGNs}

\begin{table}
\begin{minipage}{90mm}
\begin{center}
\caption{Properties of $\log \eta\prime_{\rm AGN}$ parameters for AGNs.}

\begin{tabular}{lcc}
   &   log $\eta\prime_{\rm AGN}$ (O3O4)    &   log $\eta\prime_{\rm AGN2}$ (S34O4)  \\
\hline
$N$ &   90  &  476  \\
$m$   ($\sigma$)  &  0.91  ( 0.32)  &  0.47 (0.48)  \\
\hline\\[-0.3cm]
\multicolumn{3}{c}{GMM Dist. 1}  \\
$N_1$    &   21   &   154   \\
$m_1$  ($\sigma_1$)  &   0.63 (0.31)  &   -0.06  (0.20)  \\
p-value &  0.11  &  7e-5  \\
\hline\\[-0.3cm]
\multicolumn{3}{c}{GMM Dist. 2}  \\
$N_2$   &   {\bf 69}   &  322  \\
$m_2$  ($\sigma_2$)  &  1.05 (0.23) &  0.70 (0.38)  \\
p-value &   0.014  &  3e-9  \\
\hline
\label{disp_eta}
\end{tabular}
\end{center}
\end{minipage}
\tablefoot{Number of objects, mean, and standard deviation of the log $\eta\prime_{\rm AGN}$ parameters defined for AGNs, calculated using both the [\oiii] emission line and the flux of [\siii]+[\siv]. In addition, results from a GMM for a bimodal distribution are also shown.}
\end{table}

\begin{table*}
\begin{minipage}{180mm}
\begin{center}
\caption{$\alpha_{\rm OX}$ and log $U$ obtained with {\sc HCm-Teff-IR}.}

\begin{tabular}{lcc|ccc|ccc}
\multicolumn{3}{l}{Models}  &  \multicolumn{3}{c}{$\alpha_{\rm OX}$} & \multicolumn{3}{c}{log $U$} \\
$f_e$ (\%)  &  Dust &  Lines &   $m$ ($\sigma$)  &  $m_1$  ($\sigma_1$)   &   $m_2$  ($\sigma_2$)   &   $m$ ($\sigma$)    &    $m_1$   ($\sigma_1$)  &   $m_2$   ($\sigma_2$)  \\
\hline
99.9   &  Yes &  [\siii]+[\siv]  & -1.61 (0.22)    &   -1.78 (0.07)  &  -1.36 (0.11)  &  -1.98 (0.53)   &  -1.56 (0.20) &   -2.56 (0.17)   \\
99.9   &  Yes &  [\oiii]  &  -1.54 (0.15)   &  -1.60 (0.11)   &  -1.34  (0.10)  &  -2.33 (0.30)  &  -2.04 (0.21) &  -2.50 (0.19)   \\
99.9   &  No &  [\siii]+[\siv]  & -1.53 (0.17)  &  -1.63 (0.10)  &  -1.33 (0.07)  &  -1.96 (0.53)  &  -1.52  (0.22)  &  -2.48  (0.25)  \\
99.9   &  No &  [\oiii]  &  -1.56 (0.21)  &  --  &  --  & -2.18 (0.36)  &  -1.90  (0.22)  &   -2.47 (0.20)  \\
98   &  Yes &  [\siii]+[\siv]  &  -1.52 (0.26)   &  -1.73 (0.08)  &  -1.23 (0.10)  &   -2.03 (0.54)   &  -1.61 (0.25)  &  -2.62 (0.12)  \\
98   &  Yes &  [\oiii]  &  -1.50 (0.19)  &  -1.69 (0.04)  &  -1.41 (0.16)  &  -2.30 (0.25)   &   -2.13 (0.20)  & -2.47  (0.15)  \\
98   &  No &  [\siii]+[\siv]  &  -1.50 (0.20)  & -1.62 (0.11)  &  -1.28 (0.11)  & -1.92 (0.51)  &  -1.50  (0.19)  &  -2.43 (0.26)  \\
98   &  No &  [\oiii]  &  -1.51 (0.23)   &   -- &  -- &  -2.19 (0.33)  &  -1.94 (0.23)  &  -2.42 (0.22)  \\
2   &  Yes &  [\siii]+[\siv]  &   -1.44 (0.21)  &  -1.62 (0.09)  &  -1.22 (0.09)  & -1.81 (0.42)  & -1.48 (0.17) &  -2.25 (0.18)  \\
2   &  Yes &  [\oiii]  &  -1.50 (0.25)  & -1.66 (0.11)  &  -1.22 (0.13)  &  -2.12 (0.29)  & -1.78 (0.20)  &  -2.25 (0.20)  \\
2   &  No &  [\siii]+[\siv]  &  -1.44 (0.21)   & -1.60 (0.10)  &  -1.22 (0.09)  & -1.72 (0.40)  &  -1.42 (0.15)  & -2.11 (0.25)   \\
2   &  No &  [\oiii]  & -1.47 (0.28)  &  --  &  -- &  -2.07 (0.31)  & -1.85 (0.18)  &  -2.33 (0.21)  \\
\hline
\label{disp}
\end{tabular}
\end{center}
\end{minipage}
\tablefoot{Mean $\alpha_{\rm OX}$ and log $U$ with their corresponding standard deviations of the distributions calculated by {\sc HCm-Teff-IR} for the studied sample under different assumed conditions in the models. Besides the values for the overall distribution ($m$), two values ($m_1$ and $m_2$) are given assuming a bimodal distribution, adopting a GMM, and applying a Shapiro-Wilk test so that a p-value lower than 5\% for each component confirms the bimodal hypothesis.}
\end{table*}

\begin{figure}
   \centering
\includegraphics[width=\columnwidth,clip=]{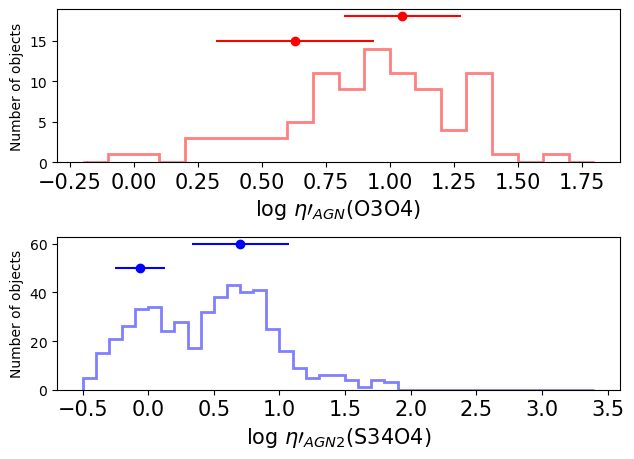}
\caption{Distribution of the derived values for log $\eta\prime_{\rm AGN}$  (O3O4; upper panel) and log $\eta\prime_{\rm AGN2}$ (S34O4; lower panel). The mean and standard deviations of the corresponding analyzed bimodal distributions are also shown.}\label{hist_eta}%
\end{figure}

First, we calculated the $\eta\prime$ parameters as defined for AGNs in Eqs. 2 and 3 for the compiled sample. The corresponding distributions for these values are represented in the histograms shown in Fig.~\ref{hist_eta}. We note that for these calculations we chose the [\oiii] line at 88 \micron\ instead of that at 52 \micron, as this allowed us to calculate the parameter in a larger number of objects in our sample. A similar consideration was taken for the version of the parameter shown in the lower panel, for which we used the [\siii] line at 18.7 \micron\ instead of that at 33.5 \micron. Similarly, we used the [\nev] line at 14.3\,\micron.

The number of points and the corresponding mean and standard deviation values for both distributions are listed in Table \ref{disp_eta}. Remarkably, a bimodal pattern for the two $\eta\prime_{\rm AGN}$ is seen in Fig.~\ref{hist_eta}, particularly in the $\eta\prime_{\rm AGN2}$ due to the higher statistical significance.
This bimodality could be partially due to the evident difference in slope between the two variable axes shown in Fig.~\ref{eta-agn} for the objects in the studied sample, being much flatter for Ne23Ne5 than for O3O4, and especially for S34O4, in contrast to the required constant slope of the different values of log $\eta\prime_{\rm AGN}$ in the same diagram. As a comparison, we show in this figure the mean values found from our analysis of the separated distributions, both for log $\eta\prime_{\rm AGN}$ and log $\eta\prime_{\rm AGN2}$.

To verify the hypothesis that the studied points can indeed be represented by a double-peaked distribution, we applied a Gaussian mixture model (GMM), assigning each point to a different component. Afterward, to confirm the bimodality found by the GMM, we applied a Shapiro-Wilk normality test independently to each of the two resulting distributions. In each case, the null hypothesis is that the objects do not correspond to a single Gaussian distribution, which is rejected if the resulting p-value is below a 5\% threshold. We note that the uncertainties in the line ratios were not taken into account for the normality test. Nevertheless, this would have reduce the weight of uncertain measurements, which typically show larger deviations.

For both representations of the $\eta\prime_{\rm AGN}$ parameter (i.e., using either [\oiii] or [\siii]+[\siv]), we verified that all p-values were below the rejection limit, confirming that each one of the two independent distributions found by the GMM are normal and hence that the whole resulting distribution is indeed bimodal.
While the overlap region between the two Gaussian components found by the GMM introduces some uncertainty in individual classifications, the statistical trends remain robust as the fraction of galaxies with ambiguous membership is small and does not significantly impact the results.

We also show in Table \ref{disp_eta} the values obtained from this analysis with the mean and standard deviation resulting for each one of the two distributions. The mean values are also shown in Fig.~\ref{eta-agn}. This result could thus indicate that two families of AGNs with a different hardness could co-exist. Nevertheless, given the dependencies shown by both $\eta\prime$ parameters for AGNs (i.e., on $\alpha_{\rm OX}$ and $U$), a more precise analysis is required to confirm this scenario.

\subsection{Results from HCm-Teff-IR}

We applied the {\sc HCm-Teff-IR} code to our compiled sample of AGNs with measured emission lines in the mid and far-IR spectral range, in order to break the possible degeneracy of the $\eta\prime_{AGN}$ parameter. In particular, we aim to explore whether the observed bimodality is still recovered when we translate the information from the emission lines into the derived parameters $\alpha_{\rm OX}$ and log $U$, the former being a parametrization of the hardness of the incident radiation field in AGNs and the latter representing the excitation state of the gas.

The use of the code also allows for a consistent comparison of results for objects presenting various emission-line combinations across different spectral ranges, thereby increasing the number of objects in the analysis. In this regard, the code admits as input any of the transition lines from [\nev], [\siii], or [\oiii], as described in Sect.~\ref{HCmTeff}. The only condition to provide a solution is that the Ne23Ne5 emission-line ratio, along with either O3O4 or S34O4, is available.\footnote{Additionally, the code admits the emission-line ratio Ar23Ar5, although this is not available for any of the objects compiled for this work.}

In Table \ref{disp} we present the results for the objects for which the code finds a solution for $\alpha_{\rm OX}$, explored in the range from -2.0 to -1.0 and log $U$ from -4.0 to -1.0. The table shows the mean values and the standard deviations of the corresponding distributions under different assumptions adopted for the models, including the change in the selected stopping criterion. This value is varied from 2\% of free electrons, which in practice implies a totally radiation-bounded geometry, to values 98\% and 99.9\%, which are those for which most of the sample is covered in the corresponding softness diagrams, as seen in Fig.~\ref{eta-agn}. Results for models computed with or without dust grains mixed with the gas, for each value of $f_e$, are also given.
We did not find any noticeable difference neither in the results nor in their uncertainties for those objects for which a previous determination of the metal content was used by the code, underlining that metallicity is not a crucial factor to derive the hardness or the excitation using IR lines for AGNs.

\begin{figure*}[t!]
  \centering
  \includegraphics[width=0.5\textwidth]{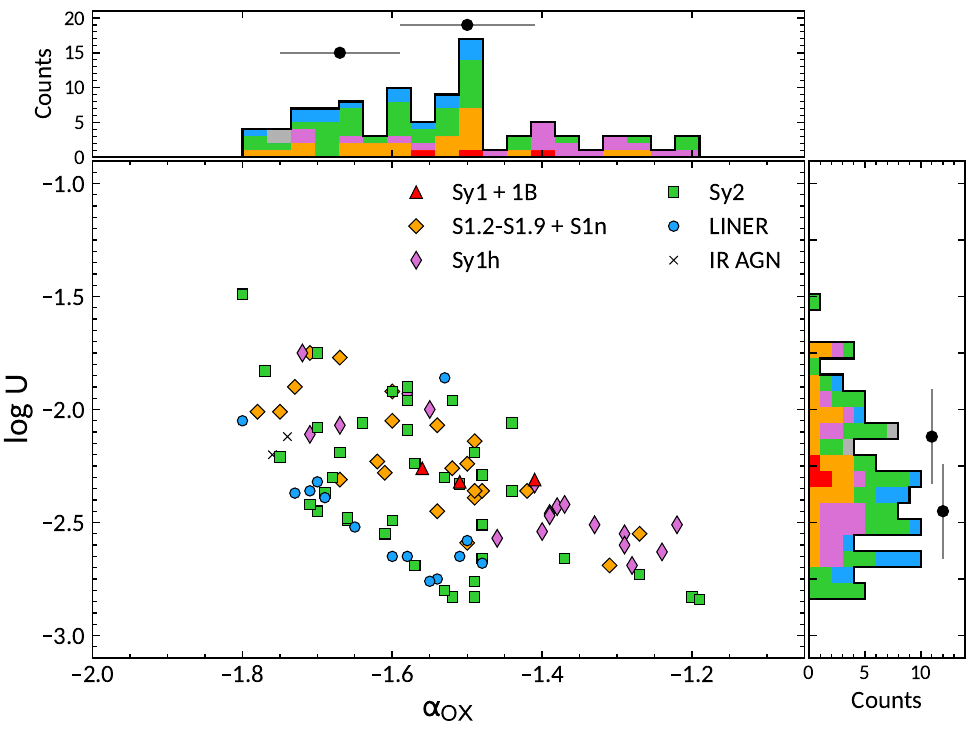}~
  \includegraphics[width=0.5\textwidth]{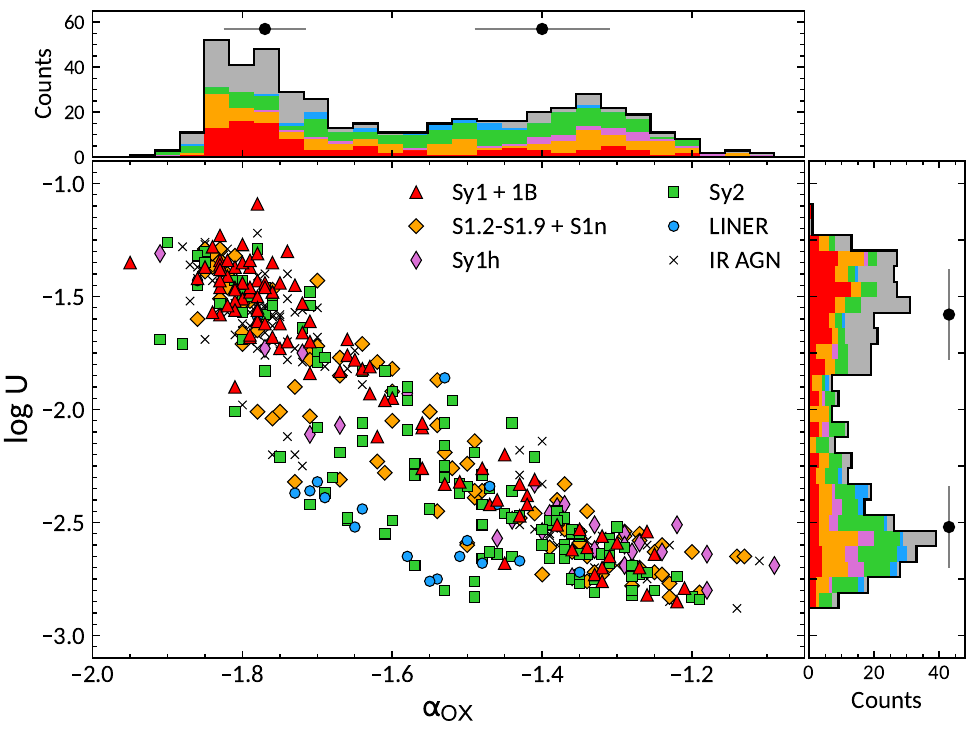}
  \caption{Scatter density plots and corresponding histograms of the resulting values for $\alpha_{\rm OX}$ and log $U$ as obtained from {\sc HCm-Teff} assuming a stopping criterion of 99.9\% of free electrons and with dust grains for the studied sample. The left panel shows those galaxies where the code used [\oiii], while the right panel shows the results using [\siii]+[\siv]. The different symbols correspond to type 1 Seyferts or optical Seyferts (Sy1 or 1B in \citealt{Spoon_2022}; red triangles), intermediate type Seyferts (Sy1.2 to Sy1.9; orange diamonds), hidden broad-line Seyferts (Sy1h; purple thin diamonds), type 2 Seyferts (Sy2; green squares), LINERs (blue circles), and IR-bright AGNs (1A, 2A, 2B, 3B in \citealt{Spoon_2022}; grey crosses). The histograms also show the marginal distributions together with the mean values obtained in the resulting bimodal distributions.}\label{alpha-U}%
\end{figure*}

From an inspection of these average results, it can be seen that the inclusion of dust in the models does not result in a noticeable difference in the outcomes. This is not unexpected, given the very small difference between the model predictions shown by the sequences in  Fig.~\ref{eta-agn}.
On the contrary, the assumption of different stopping $f_e$ values implies that $\alpha_{\rm OX}$ tends to be higher, on average, for decreasing $f_e$. This is consistent with the fact that models with $f_e$ = 2\% tend to have higher $\eta\prime$ values. However, as discussed in previous sections, this implies that a non-negligible fraction of the studied sample lies outside the model grid, so it is preferable to use the values derived from the code when matter-bounded geometry is assumed, as suggested by previous works \citep[e.g.,][]{melendez11,hcm-agn}.

Regarding the differences obtained for the two studied sub-samples (i.e., one for the objects with [\oiii] and another for [\siii]+[\siv], which it is essentially the IDEOS sample), there are noticeable differences, particularly for the $U$ values, which are significantly higher in the case of IDEOS, with averages lower than -2.0. Nevertheless, Fig.~\ref{alpha-U} shows that the overall ranges in $\alpha_{OX}$ and log $U$ values derived are similar in both cases. Thus, the discrepancy observed could be ascribed to the different sample selection criteria in the two cases. In this regard, objects with [\oiii] detections from \textit{Herschel}/PACS were likely selected among the brightest nearest Seyfert nuclei to guarantee the detection of the faintest far-IR lines. A selection bias toward more luminous objects is likely associated with higher accretion rates.

In addition to the average values, it is also important to analyze the shapes of the parameter distributions to probe how the bimodality found in $\eta\prime$ is interpreted by the code. Figure~\ref{alpha-U} shows the results obtained for the models with dust grains assuming $f_e$ = 99.9\%. The left panel shows the subsample using the O3O4 ratio, either with [\oiii] 52\micron\ or 88\micron, while the right panel corresponds to S34O4, using either [\siii] at 18\micron\ or 33\micron. The different symbols in the $\alpha_{\rm OX}$ vs. log $U$ diagram correspond to the optical or IR spectral classes indicated in \citet{veron2010} and \citet{Spoon_2022}, respectively. See Sect.~\ref{interp} for a discussion on the distribution of different AGN types on this diagram. As in the previous analysis for $\eta\prime$, the number of objects in this second diagram is much larger (480 against 97), as it involves most of the galaxies in the IDEOS catalog. We note that the total number of objects with a successful determination of $\alpha_{\rm OX}$ and log $U$ is 484 (Table~\ref{class}), which differs from the number of $\eta\prime$ values in Table~\ref{disp_eta}. While $\eta\prime$ is defined only for the 88\,$\mu$m [\oiii] line (or alternatively the 18.7\,$\mu$m [\siii] line), the code provides a solution for all the objects with detected [\neii], [\neiii], [\nev] (14.3 or 24.3\,$\mu$m), [\oiv] and either [\oiii] (52 or 88\,$\mu$m), or [\siv] and [\siii] (18.7 or 33.5\,$\mu$m). The most important aspect, however, is that the bimodal distribution of the points remains for the two variables derived by the code, that is, for both $\alpha_{\rm OX}$ and log $U$. Besides, the bimodal nature of the two variables appears to be correlated, such that objects with a higher $\alpha_{\rm OX}$ exhibit lower log $U$ and vice versa.

To probe the bimodal nature of these distributions, we performed an identical statistical treatment as that employed for the analysis of $\eta\prime$. This consisted of the application of a GMM to the results, followed by a Shapiro-Wilk normality test for each of the two resulting distributions that constitute the bimodality. The non-normality for each is rejected when the p-value, representing the null hypothesis that the distribution is not normal, is lower than 5\%. In Table \ref{disp} we provide, for those cases where the non-Gaussian hypothesis is rejected for each one of the unimodal distributions, the corresponding mean and standard deviations for each one of the two separated distributions under the different model assumptions considered by the code. As it can be seen, the bimodality is not only accepted in the case of the $\alpha_{OX}$ distributions obtained for models without dust grains. On the contrary, it is well established for the rest of cases. Thus, regardless of the specific $\alpha_{\rm OX}$ values derived with HCm-Teff-IR, which may depend on the functional form assumed for the AGN models, the bimodality observed is a robust result, as it was already present in the distribution of $\eta\prime_{\rm AGN}$.

Moreover, models help us to realize that the bimodality simultaneously appears for both the radiation field hardness and the excitation. Indeed, the correlation coefficient between these two variables is, on average across the different model assumptions, -0.95. In addition, this correlation can be easily confirmed by quantifying the relative number of objects simultaneously showing a higher $\alpha_{\rm OX}$ -- hard radiation (HR) -- and a lower log $U$ -- low excitation (LE) -- or vice versa, that is, those with a soft radiation field (SR) and a high excitation (HE), which are the combinations to most of the galaxies apparently belong. As derived from the values shown in Table~\ref{class}, this calculation yields that a 91.5\% are SRHE or HRLE, confirming that objects with a harder radiation field also have a lower proportion of ionizing photons compared to the number of particles in the gas phase. Taking the weighted mean values obtained in Table \ref{disp}, we can thus define two main AGN families: one with average $\alpha_{\rm OX}$ $\simeq$ -1.7 and log $U$ $\simeq$ -1.5 (SRHE), and another with $\alpha_{\rm OX}$ $\simeq$ -1.3 and log $U$ $\simeq$ -2.4 (HRLE).

\begin{figure*}[t!]
  \centering
  \includegraphics[width=0.5\textwidth]{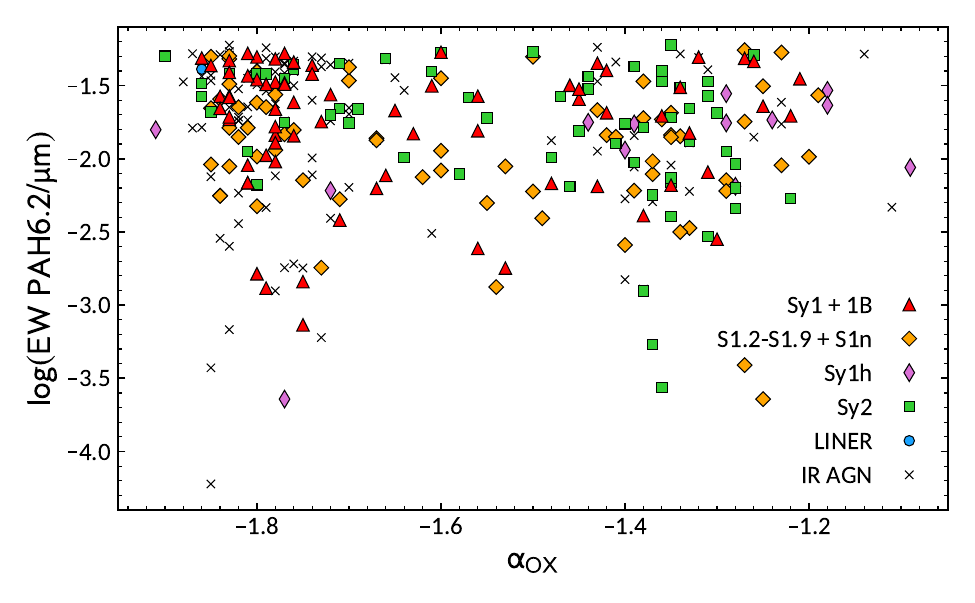}~
  \includegraphics[width=0.5\textwidth]{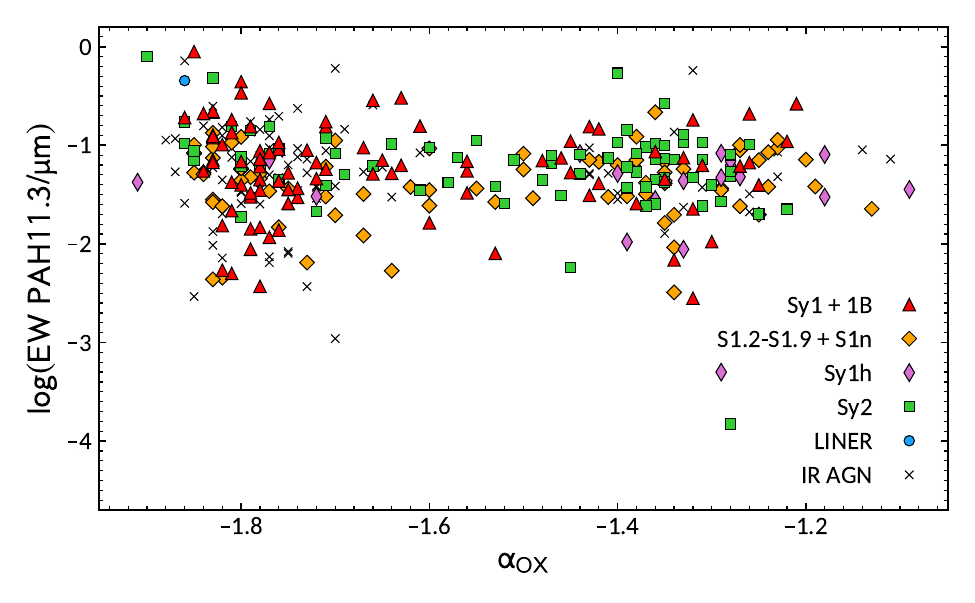}
  \includegraphics[width=0.5\textwidth]{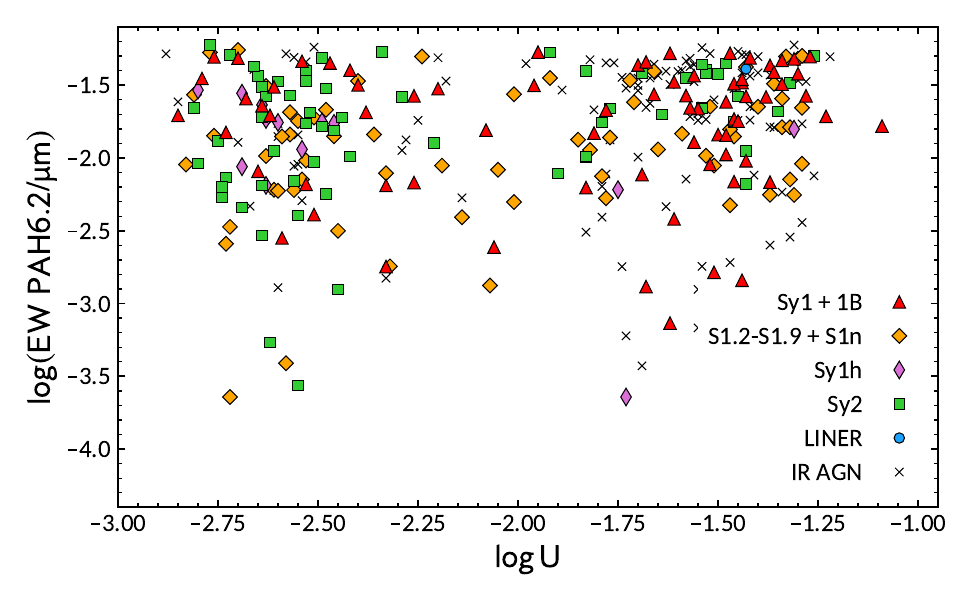}~
  \includegraphics[width=0.5\textwidth]{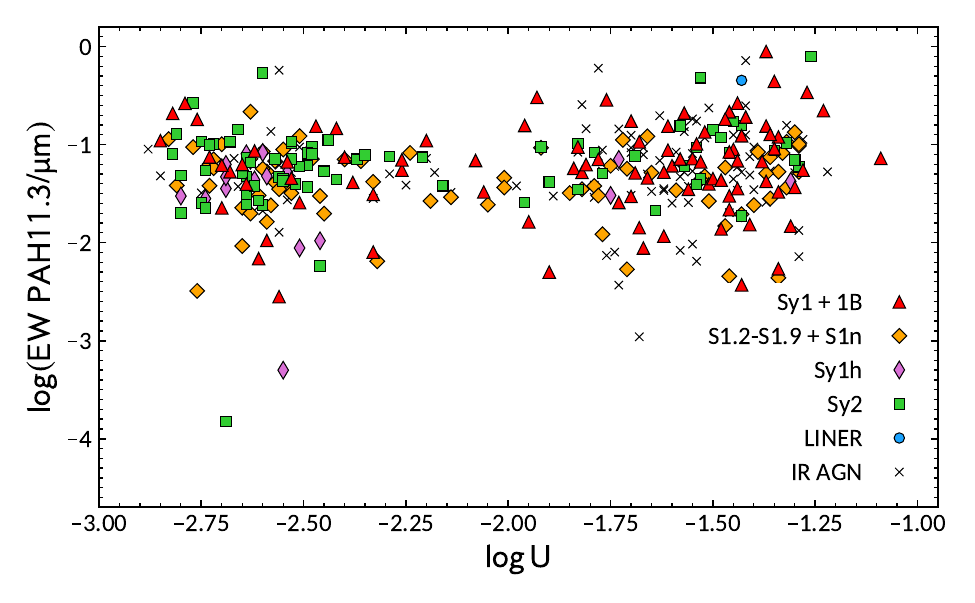}
  \caption{Relations between the derived $\alpha_{\rm OX}$ and log $U$ using the {\sc HCm-Teff-IR} code while assuming a stopping criterion for the models of 99.9\% free electrons fraction and with dust grains for the studied sample, and the measured EW of the PAHs at 6.2\micron\ and 11.3 \micron. The different symbols correspond to type 1 Seyferts or optical Seyferts (Sy1 or 1B in \citealt{Spoon_2022}; red triangles), intermediate type Seyferts (Sy1.2 to Sy1.9; orange diamonds), hidden broad-line Seyferts (Sy1h; purple thin diamonds), type 2 Seyferts (Sy2; green squares), LINERs (blue circles), and IR-bright AGNs (1A, 2A, 2B, 3B in \citealt{Spoon_2022}; grey crosses).}\label{pah}
\end{figure*}

\subsection{An interpretation of the bimodal distributions}\label{interp}

The results described above clearly indicate the existence of two well-defined families of AGNs in the Universe, based on the hardness of their radiation field and the excitation of the gas in the NLR surrounding the active nuclei. These two families emerge even when a single softness parameter is analyzed, thanks to the high statistical significance of the IDEOS sample and the use of emission lines covered by {\it Spitzer}. Moreover, the families are better characterized using the {\sc HCm-Teff-IR} code, which derives parameters uniquely tracing the hardness of the radiation field, such as $\alpha_{\rm OX}$, while simultaneously providing parameters that trace gas excitation, such as log $U$.

In the framework of AGN classification, the spectral types are commonly divided into type 1 and type 2 nuclei, as described by the unified model. Type 1 AGNs are characterized by broad emission lines (e.g., H$\beta$ and Mg\,\textsc{ii}) in their optical and ultraviolet spectra, which arise from high-velocity gas in the broad-line region (BLR) near the central supermassive black hole. In contrast, type 2 AGNs exhibit only narrow emission lines, as the BLR is obscured by dust when viewed from certain angles. This geometric interpretation has been a cornerstone of AGN taxonomy, providing a unifying framework to explain the different observed properties of these sources. However, a geometric interpretation of the spectral properties does not fully account for the physical differences in radiation field hardness and gas excitation.

Our findings challenge the traditional classification schemes, suggesting the presence of two distinct physical states in AGNs, beyond the type 1/2 dichotomy. One family exhibits hard radiation fields combined with low-excitation gas (HRLE), while the other shows soft radiation fields and highly excited gas (SRHE). These two families are robustly identified and do not depend on the assumed physical conditions in the models. By examining mid- and far-IR emission lines, we aim to elucidate the underlying mechanisms responsible for the existence of these two AGN families.

On the one hand, one may question to what extent these results can be affected by the contribution of possible star-formation emission, a critical point given the low spatial resolution provided by {\it Spitzer}. 
Taking advantage of the extensive database of spectral measurements available in the IDEOS catalog, we {investigated a possible dependence of $\alpha_{\rm OX}$ or log $U$ on the contribution from star formation activity. For this purpose we used the equivalent widths of certain PAH bands observed in the mid-IR, such as those at 6.2 \micron\ and 11.3 \micron. Although the strong radiation field in the vicinity of AGNs is expected to destroy the complex molecules responsible for these emission bands (e.g., \citealt{smith07, ds10}), a residual contribution could still be present. Thus, 
we explored possible correlations in the studied sample between the observed EWs of the PAH bands and the parameters derived by our code {\sc HCm-Teff-IR}, shown in Fig.~\ref{pah},
assuming models with dust grains and a stopping criterion of $f_e = 99.9\%$. In any case, the results obtained under different conditions are practically the same, meaning that the assumptions made in the models do not alter our results. No clear correlation is obtained for any of the four relations. A very slight, but not significant, correlation coefficient ($\rho_S$) is found for the relation between $\alpha_{\rm OX}$ and the EW of the two bands (around -0.20), indicating a lower EW value for higher $\alpha_{\rm OX}$, but this is far from conclusive. Furthermore, this correlation is mainly observed for objects belonging to the family with a lower $\alpha_{\rm OX}$ (i.e., around -1.7), while no trace of correlation is detected for objects with higher $\alpha_{\rm OX}$. Additionally, no correlation is observed for $U$, as shown in Fig.~\ref{pah}. A similar result is obtained when the EWs are represented against the $\eta\prime_{\rm AGN}$ parameters. A large dispersion is also observed in subsamples grouped by AGN type, which show no apparent correlation.

Given the lack of correlation between the hardness of the radiation field in AGNs and the presence of PAHs, we conclude that $\alpha_{\rm OX}$ or log $U$ are not affected by the contribution of star formation in our sample. Additionally, it is reasonable to consider that the spatial resolution of {\it Spitzer/IRS} observations may be insufficient, for most galaxies, to resolve the regions where PAHs are destroyed by AGN radiation from those where UV photons escaping from central galactic zones excite the vibrational modes of these molecules, thereby increasing the intensity of the PAH bands.

Nonetheless, although the low resolution could explain the lack of correlation between the hardness of the radiation field and the destruction of PAHs, it is difficult to imagine how the star-formation contribution could manifest in a clear bimodal distribution, as evidenced by both $\alpha_{\rm OX}$ and log $U$. On the other hand, the lack of correlation with the equivalent widths of the PAHs suggests that star formation is not playing a relevant role in the ionization of the observed emission lines. This is reinforced by the large [\nev]/[\neii] ratios measured in our sample, which exceed 0.2 for most of the nuclei, suggesting significant AGN contributions \citep{tom10}.

Therefore, the causes behind the observed bimodal distribution in the AGN radiation field should be sought in the AGN's own properties. In this way, our result can be interpreted as evidence that the slope of the ionizing continuum in AGNs --\,and therefore the SED shape\,-- does not exhibit a flat or smooth transition from soft to hard continua. Instead, it shows a dual-mode behavior, where active nuclei have either low- or high-hardness properties. This dichotomy was noted in previous studies of radio galaxy samples \citep{laing94,buttiglione10}. However, our analysis extends this finding to the broader AGN population. The bimodal behavior found in this study aligns with the two main accretion states identified in AGNs \citep{koerding06,svoboda17,jafo21a}, suggesting that the ionizing continuum in high-excitation sources is powered by the accretion disk, which dominates the energy output in accreting black holes during the so-called soft state or radiatively efficient mode. Conversely, low-excitation sources would be associated with the hard state or kinetic mode, characterized by a cooler and/or truncated disk, which may lead to a non-thermal continuum dominated by the jet or corona \citep[e.g.,][]{capetti05,jafo23}. The observed bimodality in both $\eta\prime_{\rm AGN}$ and $\alpha_{\rm OX}$ could then be attributed to the rapid transition between these two main accretion states.

In addition to the hardness bimodality, it is worth exploring the reasons behind the simultaneous bimodality in the NLR gas excitation and, most importantly, why the latter shows strong statistical significance in its correlation with the radiation field hardness. Specifically, most HR objects exhibit LE, while most SR objects have HE.
\begin{table}
\centering
\setlength{\tabcolsep}{3.pt}
\begin{minipage}{90mm}
\begin{center}
\caption{Object classification as a function of both $\alpha_{\rm OX}$ (i.e., SR or HR) and log $U$ (LE or HE) from {\sc HCm-Teff-IR}.}

\begin{tabular}{lcccccc}
\bf Type & \bf Total & \bf SRHE & \bf SRLE & \bf HRHE & \bf HRLE & \bf $\log\left(\frac{L^\text{[\nev]}_\text{bol}}{\rm{erg\,s^{-1}}}\right)$\\[0.1cm]
\hline\\
All   &   484   & \bf 220   &   25   &  22 &  \bf 217 & 44.91 \\
\hline\\[-0.1cm]
\multicolumn{7}{c}{\it Optical classification} \\
\hline\\[-0.3cm]
Sy1   &  87  & \bf 52 &  0 &  4 &     31 & 45.12 \\
Sy1.x &  98  &     39 &  2 &  6 & \bf 51 & 44.84 \\
Sy1h  &  30  &      3 &  2 &  2 & \bf 23 & 44.71 \\
Sy2   & 117  &     28 & 11 &  5 & \bf 73 & 44.78 \\
LINER &  21  &      3 &  6 &  1 & \bf 11 & 43.13 \\
\hline\\[-0.1cm]
\multicolumn{7}{c}{{\it IR classification}}\\
\hline\\[-0.3cm]
1B    &   32 & \bf 26  & 0 &  1 &     5 & 45.58 \\
1A    &  333 & \bf 173 & 4 & 13 &   143 & 44.88 \\
2A-3B &   16 &       6 & 0 &  2 & \bf 8 & 45.18 \\
\hline
\label{class}
\end{tabular}
\end{center}
\end{minipage}
\tablefoot{Number of objects classified in the found distributions as a function of both $\alpha_{\rm OX}$ (i.e., SR or HR) and log $U$ (LE or HE) as calculated by {\sc HCm-Teff-IR} assuming models with $f_e$ = 99.9\% including dust grains. The table includes the overall sample, and different categories as a function of their optical classification and the classes defined by \cite{Spoon_2022} from their IR spectrum (i.e., 1A: hot-dust-dominated spectrum typical of AGNs, 1B: Optical Seyferts, and 2A-3B: ULIRG mainly in interaction). For all spectral types, either the SRHE or HRLE class always contains the largest number of objects (indicated in boldface). The last column show the median bolometric luminosity for the different AGN type subsamples estimated from the [\nev] 14.3\,$\mu$m line derived by \citet{Spinoglio_2024}.}
\end{table}

To further investigate this effect, we examined the spectral classification of the galaxies in our sample. Although these are not available for all galaxies, they may reveal certain trends. Table~\ref{class} provides the number of galaxies assigned to different spectral main IR categories as defined by \cite{Spoon_2022}: 1A have hot-dust-dominated spectra typical of AGNs; 1B have IR spectra similar to those found in optical Seyferts; and 2A-3B are ULIRG, mainly in interacting systems. Additionally, the table shows the galaxy distribution across the four resulting groups formed by the combinations of the two marginal distributions for $\alpha_{\rm OX}$ and log $U$ (i.e., SRLE, SRHE, HRLE, and HRHE).  These classifications were based on models with $f_e$ = 99.9\% including dust grains, though the results are not significantly different when other model conditions are assumed. From this classification, it seems that no clear trend is found for IR-bright galaxies, such as 1A or 2A-3B types, which are more evenly distributed between the two main categories. On the other hand, it could seem that 1B galaxies, that includes Seyfert as identified from their IR spectra, are mainly classified as SRHE. Table~\ref{class} also includes the median luminosity for each AGN type subsample shown in Fig.~\ref{alpha-U}. Seyfert 1 and 1B nuclei show the brightest luminosities ($\gtrsim 10^{45}\, \rm{erg\,s^{-1}}$), although the other spectral types show similar values $\sim 10^{44.7-45.2}\, \rm{erg\,s^{-1}}$, except for LINERs, which are clearly fainter sources ($\sim 10^{43.1}\, \rm{erg\,s^{-1}}$).

This trend is partially confirmed with the results coming from the independent optical classification, also shown in Table \ref{class} and available in the objects compiled by \cite{jafo16}, and complemented for a substantial number of the objects of the IDEOS catalog in \cite{veron2010}. This classification points to a larger number of SRHE in the Sy1 type, while intermediate types (S1.x) are more equally distributed between SRHE and HRLE. On the other hand, a larger number of HRLE are found for Sy2, hidden broad-line Seyferts (Sy1h) and LINERs. Nevertheless, a non-negligible number of LINERs lie in one of the intermediate categories, such as SRLE, which could indicate that they can be more easily found in a transition process. However, the scarce number of objects classified in this category prevents us from extracting more robust conclusions. Thus, despite the apparently contradictory results for Sy2 from the optical and IR classifications, the statistical trend to find with a larger probability AGNs with prominent broad emission lines in the SRHE phase underlines the existence of a physical basis related with the variations of the accretion rate in the innermost disk, though, purely phenomenological classifications based solely on certain optical or IR spectral features are not necessarily representative of the accretion state of the active nucleus.

Moreover, the trend to identify LINERs as HRLE is consistent with several studies (e.g., \citealt{Ho99,maoz07,vagnetti13}) based on direct X-ray observations, which suggest that LINERs have a weaker blue bump luminosity combined with a less efficient radiative rate in their accretion disks (e.g., \citealt{Gliozzi2008}). This is consistent with the lower median bolometric luminosity found among the LINER subsample using the [\nev] 14.3\,$\mu$m line (see Table~\ref{class}). As a result, typical values of $\alpha_{\rm OX}$ for LINERs lie between -1.2 and -1.6, while values around -1.5 and -1.8 have been reported for Seyfert galaxies (e.g., \citealt{Constantin2009,Eracleous2010,Younes2011,Hernandez-Garcia2013}), in agreement with our results obtained from the IR classification and for a substantial fraction of the sample as compared with the optical classification.

The different causes proposed to explain the lower gas excitation observed in LINERs include the presence of high-velocity shocks, that can enhance low-excitation emission lines (e.g., \citealt{Dopita1997,Shields2003}), a possible higher gas density, which could imply a higher fraction of particles relative to the number of ionizing photons (e.g., \citealt{Filippenko1984}), or gas geometry, as a possible ionizing photon escape fraction could affect the balance between photons and particles (e.g., \citealt{Binette1996,Stasinska2008}). Nonetheless, it is difficult to establish a clear relation between the first two causes and the observed bimodality found in the analysis of mid- and far-IR lines. In contrast, the third scenario appears more plausible, considering that a sudden hardening of the radiation field within a limited-volume gas distribution can convert a radiation-bounded geometry, in which almost all photons are absorbed, into a matter-bounded one, where the most energetic photons pass through without being absorbed due to the much lower cross section of ions for high-energy photons. Thus, a change in the nature of the accretion disk could also result in a different ionization structure of the surrounding gas, leading to a much lower value for $U$ as $\alpha_{\rm OX}$ increases. In any case, more data are required, but thanks to the new ongoing and future IR facilities for low redshift (such as JWST or METIS) and for high redshift (such as ALMA), this bimodality can be explored in greater detail to determine whether a population exists that fills the gap in the bimodality observed at low redshift or if the rapid transition is confirmed.

\section{Summary and conclusions}

In this paper, we have analyzed a sample of AGN-selected galaxies with available emission-line information in the mid- and far-IR rest-frame spectral regimes in order to use adequate emission-line ratios involving high-excitation lines, such as [\nev] or [\oiv], to trace the shape of the hardness of the incident radiation field. We defined two forms of this parameter using the ratio of Ne23Ne5 to O3O4 and adapting it for a broader use by including a parameter exclusively based on mid-IR lines, S34O4. Compared to large grids of photoionization models, the newly defined $\eta\prime_{\rm AGN}$ parameters are sensitive to both the $\alpha_{OX}$, which traces the slope of the AGN continuum between the X-ray and UV ranges, and log $U$, an indicator of the gas excitation. Our study of these new softness parameters designed for AGNs revealed the existence of a clear bimodal distribution in our selected sample. This bimodal behavior is further confirmed by the application of the {\sc HCm-Teff-IR} code, which provides the corresponding distributions of $\alpha_{\rm OX}$ and log $U$.

Moreover, the existence of two well-defined AGN families has been established, which are concentrated around either a SR continuum (with $\alpha_{\rm OX} \sim$ -1.7) and HE gas in the NLR (with log $U \sim$ -1.5) or a HR continuum (with $\alpha_{\rm OX} \sim$ -1.4) and LE gas (with log $U \sim$ -2.4). The presence of these two categories can be identified regardless of the geometry and presence of dust grains assumed in the model grids used for comparison with the observed emission-line ratios. The existence of a bimodality in both the hardness of the radiation field, as seen in the NLR of AGNs and derived from mid- and far-IR lines, and in the ionization parameter is consistent with a dual scenario of accretion efficiency in black holes and variable geometry of the surrounding gas as a function of the SED emitted by the disk.

\begin{acknowledgements}
We thank the anonymous referee, whose very helpful and constructive comments and suggestions have helped to improve our manuscript. This work has been funded by project Estallidos8 PID2022-136598NB-C32 (Spanish Ministerio de Ciencia e Innovaci\'on).
We also acknowledge financial support from the State Agency for Research of the Spanish MCIU through the ``Center of Excellence Severo Ochoa'' award to the Instituto de Astrof\'isica de Andaluc\'ia (SEV-2017-0709).
JAFO acknowledges financial support by the Spanish Ministry of Science and Innovation (MCIN/AEI/10.13039/501100011033), by ``ERDF A way of making Europe'' and by ``European Union NextGenerationEU/PRTR'' through the grants PID2021-124918NB-C44 and CNS2023-145339; MCIN and the European Union -- NextGenerationEU through the Recovery and Resilience Facility project ICTS-MRR-2021-03-CEFCA.
EPM also acknowledges the assistance from his guide dog Rocko without whose daily help this work would have been much more difficult.

\end{acknowledgements}

%
   \bibliographystyle{aa} 
   \bibliography{refs.bib} 
%

\end{document}